\documentclass[11pt]{article}

\usepackage[english]{babel}
\usepackage[utf8x]{inputenc}
\usepackage{amsmath}
\usepackage{float}
\usepackage[caption = false]{subfig}
\usepackage{graphicx}
\usepackage[margin=1in]{geometry}
\usepackage{setspace}
\usepackage{hyperref}
\usepackage{cite}
\usepackage{xcolor}

\title{\large Accounting for Rink Effects in the National Hockey League's Real Time Scoring System
\vspace{-3ex}
}
\author{ Michael Schuckers\\
St. Lawrence University\\
(\href{mailto:schuckers@stlawu.edu}{\tt schuckers@stlawu.edu}) 
\and Brian Macdonald\footnote{Current address: Florida Panthers, 1 Panther Parkway, Sunrise, FL 33323}
\\ United States Military Academy}

\date{\today}
\begin{document}
\maketitle
\begin{abstract}
Recording of events in National Hockey League  rinks is done 
through the Real Time Scoring System.
This system records events such as hits,
shots, faceoffs, etc., as part of the
play-by-play files that are made publicly available.  Several previous studies have
found that there are inconsistencies 
in the recording of these events from 
rink to rink.  
In this paper, we propose a methodology for 
estimation of the rink effects for each of
the rinks in the National Hockey League.
Our aim is to build a model which accounts
for the relative differences between rinks.
We use log-linear regression to model counts
of events per game with several predictors including
team factors and average score differential.
The estimated rink effects
can be used to reweight recorded events
so that can have comparable counts of
events across rinks.  Applying our methodology 
to data from six regular seasons, we find that there are some rinks with
rink effects that are significant and consistent across
these seasons for multiple events.
\end{abstract}
\doublespacing
\section{Introduction}
In the space of four days in October 2011, the 
Columbus Blue Jackets  and the Dallas Stars 
played a home-and-home series of close games with DAL
winning both.  What is notable about these games
is that in Dallas, the Stars outshot Columbus by 16 
at even strength and in Columbus, the Blue Jackets 
outshot Dallas by 23.  This is quite a dramatic
swing in the course of a few days.  There are several possible factors that might account
for this 39 shot swing, such as home ice advantage, changes in game plan, injuries or the randomness inherent in hockey games.  Another possible reason is differences in the way events are recorded at the two arenas. 


The foundation of quality inference is quality data. 
In the National Hockey League most of the data that is publicly available comes from the  
NHL's Real Time Scoring System or RTSS, \cite{rtss}.
Unfortunately,
it has long been known that 
recording tendencies for the RTSS data in the 
National Hockey League tend to vary from rink to rink.   
This is not unexpected, since many of these statistics are subjective, and it is very 
difficult for two different scorers in two different arenas to record the data in exactly the same way.
Those differences have been covered by several authors including
Desjardins \cite{desjardins_MSG}, 
McCurdy \cite{mccurdy}, 
Fischer \cite{fischer}, 
Zona \cite{zona} and 
Awad \cite{awad}. 
Since many advanced methodologies, e.g. 
CorsiRel \cite{corsirel}, 
Fenwick \cite{fenwick}, 
Macdonald's Expected Goals  \cite{bmac_eg}, 
Schuckers and Curro's Total Hockey Ratings \cite{thor13}, and Johnson's HART 
\cite{hart}
depend on the 
RTSS data, it would be useful to account for 
those rink differences.  The goal of this project is 
to do just that:  estimate the differences in recording
tendencies at each rink in order to adjust for those effects.  

In this paper we fit statistical 
regression models to estimate 
the differences in the recording of events by rink.  We 
use 5-on-5 non-empty net even strength (ES) regular season data 
from the 2007-08 season through the 2012-13 season 
adjusted for the total amount of ES time per game.  
In addition to the rink in which a given game occurred 
we include the teams involved in the game as well as 
other explanatory covariates 
which affect the rates of each event.  The events we 
considered are the following: SHOTs, MISSs, BLOCKs, HITs, giveaways (GIVEs), takeaways (TAKEs), turnovers (TURNs), CORSIs which are 
SHOTs + GOALs + MISSs + BLOCKs, and FENWICKs which are SHOTs + GOALs + MISSs. A turnover is a giveaway or a takeaway, 
as defined by Schuckers and Curro, 
\cite{thor13}.  
The primary
result of this model is a set of multipliers for 
reweighting each event to counteract these rink effects.  

Overall we find that for the most part 
NHL rinks and arenas (and the individuals therein) do a reasonably consistent 
job of recording events.  
This is especially true of the recording of SHOTs which has the fewest rinks with significant recording issues and 
which has the smallest rink effects that we found.  Other counts are impacted by rink effects but 
ratios of events such as Corsi For Percentage remain relatively unaffected.  
The effects of factors such as `home ice' and score differential on these events are
 consistent from year to year which 
suggesting that the recording of events
has been stable over the six years of data
that we have studied.  

However, there are 
some rinks that have significant recording tendancies that make them different from the rest of the league on certain events that are recorded as part of the RTSS system.  
In particular, the rinks in Edmonton, Los Angeles, New Jersey and Toronto consistently inflate or deflate a number of the RTSS events recorded in those arenas.  To account for these we provide
rink adjustment multipliers for each event.  These adjustments are useful for correcting statistics for both teams and players playing in the affected rinks.  Since the RTSS statistics are used in contract negotiations, and since in a given year 
players play half of their games at a single
rink, there is a clear need to make these data as comparable as possible.  Using our adjustments provides improved standardization of event counts across rinks which results in more consistent player data.

\section{Data}
To build the data for estimation of the rink effects, we used the {\tt nhlscrapr} R package of Thomas and Ventura \cite{nhlscrapr} to obtain the recorded events 
for each game.  The data we are analyzing comes from six recent NHL regular seasons: 2007-08 through 2012-13.  
We found the counts for each of the events recorded during non-empty net five-on-five (NEN5v5) hockey
during the first three periods.  The primary events
we considered were BLOCKs, GIVEs, HITs, MISSs, SHOTs, TAKEs.  For determination of the relative rates of these events
at a given rink
we prorated the events in each game by dividing the number of events by the seconds of NEN5v5 time and multiplying by
3600.   For example, if there were 14 HITs for the home in 3000 seconds of NEN5v5 then the prorated total was $14/3000\times3600=16.8$.  We did
this for all events to obtain rates of events per game.  We also calculated the prorated CORSI and FENWICK 
rates which come from adding SHOTs + GOALs + MISSs + BLOCKs and SHOTs + GOALs + MISSs, respectively.  GOALs were not
prorated as they are not at the discretion of the RTSS system.  
Below we will look at the results for TURN as well.

In addition to the prorated counts for each event, we also record the team responsible for the count, the opposing team, which
team was the home team, the rink in which the events occurred, and the average score differential (ASD).  
Because it is well known that how teams play changes depending on the closeness of the game, we included a measure
of how close the given game was.  To do this we took the score differential (home score - away score) at each 
recorded event (in the first three periods) and multiplied it by the length of time until the next event.  We then divided this
total by 3600.  That is,
\begin{equation}
ASD=\frac{1}{3600} \sum_t (H_t - A_t) \ell (t)
\end{equation}
where $H_t$ and $A_t$ are the home team and the away team scores, respectively  at event $t$ and $\ell(t)$ is the length of time corresponding to that event.
Fortunately, the length of time, $\ell (t)$, for each event is included as part of the {\tt nhlscrapr} package.  
Thus our metric here is the average lead (or deficit) of the home
team over the course of the game. For example, in a game where the home team leads by 1 goal for 1240 seconds and by 2 goals for
213 seconds, the ASD here would be $0.463 = (1*1240+2*213)/3600$.  Note that for ASD we include all events not just those NEN5v5 events. 

The aim here is to be able to assess the impact on each event type that is at the discretion of the RTSS recorders at each rink.  We will call the impact of each 
arena a rink effect (RE).  
To estimate the RE's for each rink, we need to remove the impact of some other important factors from the prorated count.
Above we mentioned the impact of score which is measured via the ASD. 
In our model we will additionally include both the team responsible for the event (e.g. the team taking the shot) and the opposing
team.  The idea here being that the rate of events can be impacted by these teams.  Note that this
means that we will have two entries in our data for each game, one for each team.  One of those teams will
be the home team and we will explicitly account for the team with `home ice' in our model \footnote{For games played at non-traditional sites such as in Stockholm (NHL Premiere) or at Fenway Park (Winter Classic) we used the team that the National Hockey League designated as the home team for these games though technically they were not at home.  There were 25 of these games out of the 6858 games that we analyzed.}.
Finally, we will
%
%
have an interaction between the home team effect and the rink effect.  This term will be for the possibility that
the recording at the home rink is not the same for both teams.  We think of this effect as the `homer' effect.  

\section{Models}
Our statistical model for each event is one that
equates the natural logarithm of the prorated counts with 
a sum of an overall mean effect, a rink effect, 
an average score differential effect, a home
team effect, a `for team' effect, an `opposing
team' effect and a rink by home ice interaction
effect.  A logarithmic transformation is a common one for the modelling of rate data since it is the canonical
link for Poisson (or count) data.  See, for example, McCullagh and Nelder \cite{mccullagh_nelder}. 

Notationally, our yearly model is the following:
\begin{equation}
\label{yeareqn}
ln(Y^{*}_i) = \mu + \gamma_j + \beta ASD_i + \phi_k + \omega_{\ell} + \eta I_{j=k}  + \eta\gamma_j I_{j=k}+ \epsilon.
\end{equation}
$Y^{*}_i$ is the prorated count for team-game $i$, while
$\mu$ is the natural log of the mean effect \footnote{To deal with games in which zero events occurred we added $10^{-3}$ to each count in order to avoid undefined values for $ln(Y_i)$.}.  The main
focus of our estimation will be the $\gamma_j$'s which
are the REs, where $\gamma_j$ is the effect of rink $j$.  We represent the ASD of game $i$ by $ASD_i$ and $\beta$ is the slope of that effect. The impact of the `for team' involved
is denoted by $\phi_k$ and the opposing team denoted
by $\omega_{\ell}$.  The home ice effect and the
home ice by rink interaction are characterized
by $\eta$ and $\eta\gamma_j$, respectively.  The indicator function, $I_{j=k}$ is one if the home team is the `for team' and is zero otherwise.  The
unit of observation is the team-game and hence
we have a pair of observations, one for each team,
per game.  We represent the error term here with $\epsilon$.

The impact of the logarithmic transformation is to have a model that is multiplicative
rather than additive.  That is, we assume that the impact of each of the effects in our model
 is multiplicative rather than additive.  Consequently, when we want to adjust RTSS data based upon RE's we can reweight each event of that type dividing the event by the reciprocal of the RE for that rink. For example, the rink effect of BLOCKs in Boston is 0.866 over this period.  This means that blocks are recorded in Boston at a rate lower than the league average accounting for the other factors in our model.  To adjust for this we can reweight each BLOCK recorded in Boston by $1/0.866 =1.154$.  Note that because of our statistical model, we explicitly account for the quality of the teams and the score differential in each game to get our estimated REs. 
 
In addition to individual yearly RE's, we investigated the \textit{average} RE's for each rink and each event across all six regular seasons.  To accomplish this, we used a fuller model than found in Equation \ref{yeareqn}.  That model is:
\begin{equation}
\label{alleqn}
ln(Y^{*}_i) = \mu_t + \gamma_j + \beta_t ASD_i + \phi_{kt} + \omega_{\ell t} + \eta_t I_{j=k}  + \eta\gamma_{j} I_{j=k}+ \epsilon.
\end{equation}
The $\phi_{kt}$'s and the $\omega_{\ell t}$'s --- with changed subscripts --- now represent the estimated team for and team against effects for each season $t$.  Our model here, Equation \ref{alleqn}, has a terms for the home ice effect, $\eta_t I_{j=k}$, for each year while the home*rink interaction is assumed to be the same for all years in this model as is the rink effect.  We do this to get an average home rink effect and an average home*rink interaction effect which we will report when the effects are consistent across seasons.  

To estimate the RE's, we
 fit the models given in Equation \ref{yeareqn} and \ref{alleqn}. 
For fitting of the model,  we use an elastic net via the
{\tt glmnet} package of Friedman et al. 
\cite{friedman10} 
in R.  For each season 
we use $10$-fold cross-validation to estimate
the $\lambda$, or the amount of shrinkage, of the elastic net. 
In all of these models, we choose to use the value of $\lambda$ that minimizes the predictive standard error of the model. 
In order for us to consider a rink to have a \textit{persistent} RE for a given event, the RE for that rink had to satisfy the following conditions
 \begin{enumerate}
 \item the RE's had to be non-zero in the same direction (always positive or always negative), in five of the six yearly models (Equation \ref{yeareqn}), 
 \item all \textit{significant} yearly RE's had to have the same direction (i.e. positive or negative), and
 \item the RE's in the fuller model (Equation \ref{alleqn}) had to be significant and in the same direction as the yearly models.
 \end{enumerate}
The choice of five of six years was drawn from the fact that we observed that some rinks were not significant for the 2012-13 season which was shortened due to a lockout.  

%
%
%

 \begin{table}[ht]
\begin{center} 
\caption{ Yearly Coefficients for BLOCKs.}
\label{table:block:years}
\begin{tabular}{l|r|r|r|r|r|r}
& \multicolumn{6}{c}{Season}\\
Effect&2007-08&2008-09&2009-10&2010-11&2011-12& 2012-13\\
\hline
mean, $\exp(\hat{\mu})$	&10.776	&11.382	&12.029	&13.017	&13.061&	13.184\\
ASD, $\exp(\hat{\beta})$	&0.924&	0.934&	0.939&	0.920&	0.928&	0.931\\
home ice, $\exp(\hat{\eta})$&	1.070&	1.045&	1.051&	1.058&	1.056&	1.037
\end{tabular}
\end{center}
\end{table}
 
 \section{Results for Rink Effects}
 In this section we will discuss the results of
 fitting our models for each event type.  Overall
 there is a great deal of continuity from year to year in these results. In the sections that follow we look at the REs for BLOCKs, GIVEs, HITs, MISSs, SHOTs, TAKEs, CORSIs, FENWICKs, TURNs.  The last three in this list are derived metrics largely based upon the first six events.  CORSI events are SHOTs, GOALs, MISSs and BLOCKs.  FENWICK events are SHOTs, GOALs, and MISSs, while TURNs events are TAKEs by a 'for team' and GIVEs by the 'against team'.  TURNs were created 
 to deal with the issue of REs in the recording of TURNs and GIVEs.        
 
 \subsection{Results for BLOCK}
A BLOCK is recorded when the `for team' takes a shot that is blocked from reaching the goal by the `against team.'
The results for the BLOCK event from our model can be found in Tables \ref{table:block:years} and \ref{table:block:rinks}.  Starting with the results in Table 1, we see that the prorated average number of shots that are blocked per team has risen slightly over the six seasons covered here from about $11$ in 2007-08 to $13.2$ in 2012-13.  While the mean has been trending up, the effect of score differential had remained roughly the same  
over that same time period at around $0.928$.  
That is, for an average goal differential of 1, a team will have  $93\%$ of BLOCKs that they would if the average goal differential was $0$, zero.  
This is likely to be a reflection of the fact that teams who are leading a game generally take fewer shots and, consequently, have fewer 
shots that are blocked.  Home teams are 
roughly consistent in getting about $5\%$ more 
BLOCKs than the opposition. 
%
%
%

There are quite a few rinks where the recording of blocks is significantly different from the rest of the league.  These rinks are home to the following teams: Anaheim (ANA), Boston (BOS), Carolina (CAR), Columbus (CBJ), Detroit (DET), Edmonton (EDM), Florida (FLA), Los Angeles (L.A), Montr\'{e}al (MTL), New Jersey (N.J), New York Islanders (NYI), Philadelphia (PHI), San Jose (S.J), Toronto (TOR) and Washington (WAS). Table \ref{table:block:rinks} has the estimated REs for each of these rinks. N.J is particularly different from average as they count BLOCKs at a rate that is nearly half, about $54\%$, of the league average, while in Montr\'{e}al
 and Toronto BLOCKs are counted at a rate that is over 24\% greater than league average.  
 In the case of blocks, there is no evidence that there is a significant home team by rink interaction for any of the NHL's rinks. 

 \begin{table}[htb]
\begin{center} 
\caption{ Significant Rink Effects for BLOCKs.\label{table:block:rinks}}
\begin{tabular}{l|r}
Rink & Effect\\
\hline
MTL&1.271\\
TOR& 1.245\\
 NYI& 1.208\\
  S.J& 1.196\\
WSH& 1.187\\
CAR&1.187\\
 EDM&1.124\\
 PHI& 1.123\\
 \hline
 DET& 0.879\\
L.A& 0.866\\
FLA&0.868\\
 BOS& 0.852\\
CBJ&0.852\\
ANA & 0.721\\
N.J& 0.541\\
\end{tabular}
~\\
There were no persistent 'homer' effects for BLOCKs.
\end{center}
\end{table}

 \subsection{Results for GIVE}
 
 For the NHL's RTSS system a GIVE is a giveaway from the `for team' to the `against team'.  As seen in Table \ref{table:give:years}, there has been a good deal of fluctuation in the average number of GIVEs per season with a high of $6.3$ in 2009-10 and a low of $4.8$ in 2010-11. Score differential has a significant effect on pace adjusted recording of GIVEs with the number of GIVEs decreasing to $95\%$, on average, when the average score diffential is one.  There is not a trend across seasons for the score differential effect here.  Given what has been written about GIVE (and TAKE) previously it is not surprising that we see a large home ice effect here. 
See, for example, McCurdy \cite{mccurdy}.  
 All other things being equal we see that the home team is credited with a rate of GIVEs that is about $68\%$ higher than for the visiting team.     
 
 There were 18 rinks where the recording of GIVEs was significantly different from the league average.  These rinks were home to the following teams:
 Carolina (CAR), Columbus ( CBJ), Chicago( CHI),
 Dallas(DAL),  Edmonton(EDM),
Florida( FLA), Los Angeles ( L.A),
 Minnesota(MIN), Montr\'{e}al ( MTL),
New Jersey( N.J), New York Islanders
( NYI), New York Rangers ( NYR), Ottawa (OTT), Phoenix (
 PHX), Pittsburgh ( PIT), San Jose ( S.J), 
St.~Louis (STL) and Toronto ( TOR).  There is a great deal of variation in REs for GIVEs from Columbus where the recording of GIVEs runs at about $14\%$ of the league average rate to Edmonton where that same rate is over $200\%$ of league average.  
 As was the case with BLOCKs, there were not significant home by rink interactions for GIVEs. 
 
 \begin{table}[ht]
\begin{center} 
\caption{ Yearly Coefficients for GIVEs.\label{table:give:years}}

\begin{tabular}{l|r|r|r|r|r|r}
& \multicolumn{6}{c}{Season}\\
Effect&2007-08&2008-09&2009-10&2010-11&2011-12& 2012-13\\
\hline
mean, $\exp(\hat{\mu})$&5.007&	5.775&	6.258&	4.903&	4.768&	4.974\\
ASD, $\exp(\hat{\beta})$	&0.947&	0.966&	0.947&	0.963&	0.942&	$^{\dagger}$1.000\\
home ice, $\exp(\hat{\eta})$	&1.686	&1.685	&1.684	&1.690	&1.671&	1.671
\end{tabular}
\\
$^{\dagger}$ denotes that this coefficient was not significant
\end{center}
\end{table}

 \begin{table}[ht]
\begin{center} 
\caption{ Significant Rink Effects for GIVEs.}
\begin{tabular}{l|r}
Rink & Effect\\
\hline
 EDM&2.167\\
 L.A&1.748\\
 NYI&1.612\\
 S.J&1.560\\
DAL&1.539\\
 TOR&1.524\\
MTL&1.337\\
OTT&1.298\\
\hline
MIN&0.547\\
 CAR&0.499\\
 PIT&0.424\\
 STL&0.396\\ 
 CHI&0.293\\
 NYR&0.278\\
 N.J&0.270\\
  FLA&0.244\\
PHX&0.189\\
 CBJ&0.144\\
\end{tabular}\\
There were no persistent 'homer' effects for GIVEs
\end{center}
\end{table}

 \subsection{Results for HIT}
  Next we consider the effect of hits.  From Table \ref{table:hit:years} we can see that there has been an increase in the pace-adjusted number of hits per game over this period though that number has stabilized over the past four years.  The overall pace-adjusted rate of HITs has been around $24$ hits per 60 minutes of even strength play.  The first three years of our data have prorated HIT rates that increase from $20.8$ to $23.6$.  The rates then seem to stabilize around a little more than $24$ HITs recorded per team.  Unlike the mean effect, the score differential effect has been relatively stable in this period.  Teams that lead tend to have fewer HITs recorded than teams that trail over the course of a game.  The home ice effect is similarly stable over this period with home teams having hits recorded with a rate of about $11\%$ over visiting teams.  
  
There were twelve teams that had significant REs from our analysis.  These teams can be found in Table \ref{table:hit:rinks}.  The largest REs were in Los Angeles (L.A), $1.298$  and New York Rangers (NYR),$1.274$ while the smallest effects were found in New Jersey (N.J), $0.592$ and Calgary (CGY), $0.639$.  This means that games in New Jersey resulted in $40.8\% =100\% -59.2\%$ fewer HITs than would be expected based upon the teams involved and the ASD of the game while HITs in Los Angeles were recorded at a rate, $1.298$, that was nearly $30\%$ higher than would have been recorded at other rinks.
Note that we found a significant RE for Winnipeg/Atlanta here but we do not report it in our table since that effect was split over two arenas.  Presumably there are different people recording events in Winnipeg than were doing that same job in Atlanta. 
In addition to the significant REs, there are significant interactions between the rink effects and the home ice effects for three rinks.  These rinks are Dallas, New Jersey and Toronto.  That these interactions are significant means that HITS in those rinks are recorded differently for home teams and away teams.  In the cases of New Jersey and Dallas, the rate of HITs by the home team as having higher rates of HITs than would be expected from the home ice effect and the individual rink effects alone, while in Toronto the rate of HITs by the home team is about $10\%$ lower.  
 
 \begin{table}[ht]
\begin{center} 
\caption{ Yearly Coefficients for HITs.
\label{table:hit:years}}
\begin{tabular}{l|r|r|r|r|r|r}
& \multicolumn{6}{c}{Season}\\
Effect&2007-08&2008-09&2009-10&2010-11&2011-12& 2012-13\\
\hline
mean, $\exp(\hat{\mu})$&
20.762	&22.339&	23.647&	24.433&	24.672&	24.193\\
ASD, $\exp(\hat{\beta})$&0.978	&0.968	&0.970&	0.968&	0.965&	0.979\\
home ice, $\exp(\hat{\eta})$&1.144&	1.104&	1.117&	1.099&	1.111&	1.098
\end{tabular}
\\
$^{\dagger}$ denotes that this coefficient was not significant
\end{center}
\end{table}

\begin{table}[ht]
\begin{center} 
\caption{ Significant Rink Effects for HITS.\label{table:hit:rinks}}
\begin{tabular}{l|r}
Rink & Effect\\
\hline
L.A& 1.298\\
NYR& 1.274\\
PHX&1.163\\
DAL&1.197 \\
FLA& 1.175\\
TOR&1.132\\
OTT&1.091\\
\hline
COL&0.848\\
EDM&0.805 \\
MIN& 0.783\\
CGY&0.639\\
N.J& 0.592\\
\hline
N.J*HOME&1.196\\
DAL*HOME&1.073\\
TOR*HOME&0.906\\
\end{tabular}
\end{center}
\end{table}

\subsection{Results for MISS}
 
A MISS is a shot attempt that is neither blocked nor is on target for the net. 
As was the case with HITs the pace-adjusted rate of MISSs has increased over
the years considered here from about $9.2$ during the 2007-8 season to $10.3$ in the 2012-13 season.  These results can be seen in Table \ref{table:miss:years}.  The score differential effect is relatively stable over this same period, with each additional goal of average score differential leading to a decrease in the rate of MISSs of about 5\%.  The home ice effect seems to have decreased slightly for MISSs over the time period of this study from a rate of about 12\% higher in 2007-08 to a rate of about 3\% higher in 2012-13.  

The persistent REs for MISSs can be found in Table \ref{table:miss:rinks}.  The rinks that had significantly higher rates of MISSs were Carolina (CAR), Dallas (DAL), Edmonton (EDM), Los Angeles (L.A) and Toronto (TOR).  Rinks with significantly lower rates of MISSs were Boston (BOS), Columbus (CBJ), Chicago (CHI), and New Jersey (N.J).  Especially noteworthy is the very low rates of MISSs in Chicago and New Jersey both with rates about $60\%$ of the league average.

There are two rinks for which there is an interaction between the home ice effect and the RE.  Those rinks are Colorado (COL) and the New York Rangers (NYR).  Colorado actually counts fewer MISSs for the  home team while at the Rangers home rink they count more MISSs for the home team relative to the league average.  In both cases, those same rinks do not generally record counts above (COL) or below (NYR) the league average rate of the number of MISSs, only the number of MISSs by the home team.
 
\begin{table}[ht]
\begin{center} 
\caption{ Yearly Coefficients for MISSs.\label{table:miss:years}}

\begin{tabular}{l|r|r|r|r|r|r}
& \multicolumn{6}{c}{Season}\\
Effect&2007-08&2008-09&2009-10&2010-11&2011-12& 2012-13\\
\hline
mean, $\exp(\hat{\mu})$&
9.209	&9.747	&10.065	&10.311	&10.316	&10.291\\
ASD, $\exp(\hat{\beta}$)&0.940	&0.968	&0.946	&0.941	&0.957	&0.956\\
home ice, $\exp(\hat{\eta}$)& 1.127	&1.108	&1.109	&1.069	&1.077&	1.032\\
\end{tabular}
\end{center}
\end{table}
 \begin{table}[ht]
\begin{center} 
\caption{ Significant Rink Effects for MISSs.\label{table:miss:rinks}}
\begin{tabular}{l|r}
Rink & Effect\\
\hline
 TOR&1.250\\
CAR&1.211\\
 DAL&1.183\\
 L.A&1.167\\
EDM&1.111\\
\hline
BOS&0.880\\
CBJ&0.863\\
 N.J&0.603\\
 CHI&0.562\\
 \hline
  NYR*HOME&1.125\\
 COL*HOME&0.977
\end{tabular}
\end{center}
\end{table}

 \subsection{Results for SHOT}
 
The number of SHOTs that a team takes is an important metric.  Table \ref{table:shot:years}  has a summary of the overall trends in SHOTs for the data analyzed here.  The prorated estimated overall rate for SHOTs has increased over the six seasons of these data from about $23.6$ in the first year.  The subsequent years have a mean rate that seems to fluctuate around $25.5$.  The score differential effect seems to have dropped over this range of seasons from an effective slope of about $0.966$ in 2007-08 and 2008-09 to about $0.956$ in the more recent seasons.  Similarly home ice effects seemed to have lessened over the these six seasons.  These are slight changes.  

The number of persistent REs for SHOTs is much less than for the other events we have seen so far.  There are only two rinks, Florida (FLA) and St. Louis (STL), where the number of shots is significantly different from the overall league rate. These effects for SHOTs are relatively small compared to the persistent rates of other events such as HITs and GIVEs, for example. And there are no rinks where there is a significant home ice by rink interaction for SHOTs.

\begin{table}[ht]
\begin{center} 
\caption{ Yearly Coefficients for SHOTs.\label{table:shot:years}}

\begin{tabular}{l|r|r|r|r|r|r}
& \multicolumn{6}{c}{Season}\\
Effect&2007-08&2008-09&2009-10&2010-11&2011-12& 2012-13\\
\hline
mean, $\exp(\hat{\mu})$	&23.637	&24.935	&25.313	&26.448&25.616&	25.089\\
ASD, $\exp(\hat{\beta})$	&0.964	&0.967	&0.955	&0.957	&0.955&	0.957\\
home ice, $\exp(\hat{\gamma})$	&1.077&	1.061&	1.051&	1.054&	1.062&	1.033
\end{tabular}
\end{center}
\end{table}

 \begin{table}[ht]
\begin{center} 
\caption{ Significant Rink Effects for SHOTs.\label{table:shots:rinks}}
\begin{tabular}{l|r}
Rink & Effect\\
\hline
FLA&1.030\\
\hline
STL&0.955\\
\end{tabular}
\\
There were no persistent 'homer' effects for SHOTs.
\end{center}
\end{table}

 \subsection{Results for TAKE}
 
 The last of the raw events that we will consider here is the takeaway or TAKE.  A takeaway occurs when one team takes the puck from another.  Results for the overall effects for TAKEs can be
 found in Table \ref{table:take:years}.  Like many of the other RTSS events we have discussed there seems to some changes in the mean overall rate of TAKEs over the six seasons studied here.  In this case the rate of TAKEs increases from a low of $3.8$ in 2007-08 to a high of $5.3$ in 2011-12 but then fell back to $4.4$ in 2012-13.  Again it is important to mention that 2012-13 was not a full season and so estimates from that season have more uncertainty than other estimates.  In all years, the rate of TAKEs as the average score differential increases is never less than the mean rate.  There seems to be an upward trend in the estimated slope due to the score differential.  Meanwhile the home ice effect has a good deal of variability overall and much like the mean effect rises and falls through the seasons covered here.  However, on average the home team is credited with about $50\%$ more TAKEs than the away team. 
 As noted above, TAKEs and GIVEs have been known to have large home ice effects for some time.  See, for example, Desjardins \cite{desjardins_give}.  This is clearly confirmed by the analyses here.  
 
 In addition to the overall model effects for TAKEs, there are fourteen significant REs here.  The following rinks had REs that were significantly below the overall league rate: Boston (BOS), Buffalo (BUF), Columbus (CBJ), Florida (FLA), Los Angeles  (L.A), Minnesota (MIN), New Jersey (N.J), Phoenix (PHX), Pittsburgh (PIT) and St. Louis (STL).  On the other side, the following rinks had rates of TAKEs that were significantly above the league average rates: Carolina (CAR), Calgary (CGY), Colorado (COL), Edmonton (EDM), Nashville (NSH), New York Islanders (NYI) and Ottawa (OTT).  Pittsburgh and NYI were outliers, having recording rates that were approximately one-fifth (20\%) and twice (200\%) of the league rate, respectively, after adjusting for the other factors in our model.  
 There were no significant home ice by team interactions for TAKEs.
 
 \begin{table}[ht]
\begin{center} 
\caption{ Yearly Coefficients for TAKEs.\label{table:take:years}}

\begin{tabular}{l|r|r|r|r|r|r}
& \multicolumn{6}{c}{Season}\\
Effect&2007-08&2008-09&2009-10&2010-11&2011-12& 2012-13\\
\hline
mean, $\exp(\hat{\mu})$&	3.802	&4.816&	4.905&	5.098&	5.296&	4.365\\
ASD, $\exp(\hat{\beta}) $	&1.008&	1.026&	1.056&	1.093&	1.089&	1.049\\
home ice, $\exp(\hat{\gamma})$&	1.522	&1.626&	1.580&	1.545&	1.523&	1.432
\end{tabular}\\
$^{\dagger}$ indicates that the estimate was not significant.
\end{center}
\end{table}

 \begin{table}[ht]
\begin{center} 
\caption{ Significant Rink Effects for TAKEs.\label{table:take:rinks}}
\begin{tabular}{l|r}
Rink & Effect\\
\hline
 NYI&1.943\\
 OTT&1.483\\
 NSH&1.545\\
COL&1.540\\
 EDM&1.259\\
 CGY&1.212\\
 CAR&1.208\\
\hline
FLA&0.671\\
STL&0.636\\
 N.J&0.615\\
 CBJ&0.527\\
 BOS&0.492\\
L.A&0.487\\
MIN&0.445\\
PHX&0.430\\
 BUF&0.382\\
PIT&0.214
\end{tabular}
\\
There were no persistent 'homer' effects for TAKEs.
\end{center}
\end{table}

 \subsection{Results for CORSI}

We move now to CORSI events.  Unlike the previous events, CORSI events are derived metrics, not something that is recorded as part of the RTSS system.  Because of the interest in CORSI as a measure of team performance, we analyzed data about CORSI using our models.  As we can see from Table \ref{tab:corsi:years}, the mean rate of recorded CORSIs has increased over this period by
about ten percent from about $48$ per 60 minutes of even strength to over $52$ per 60 minutes during the six seasons under consideration here.  The effect of score differential has been relatively steady at about $0.965$ across all six seasons.  This means that for each additional average goal lead over the course of a game, the team leading will have a rate CORSIs of $96.5\%$ of what they would have had in a game with an average score differential of zero.  Finally we note that the impact of home ice has been relatively stable.  High values in 2007-08 and low values in 2012-13 bookend these effects while there is relative stability about the middle four years.  On average home teams have CORSI events that are recorded at a prorated rate that is $1.06$ times higher than away teams.  

There are twelve rinks with persistent REs for CORSI.  Rinks that signficantly count  CORSIs at a rate below the league average are Boston (BOS), Columbus (CBJ), Chicago (CHI), Florida (FLA), New Jersey (N.J) and St. Louis (STL).   Rinks that significantly count CORSIs at an above average rate are Carolina (CAR), Edmonton (EDM), Montreal (MTL), Philadelphia (PHI), San Jose (S.J), and Toronto (TOR).  There were no signficant interactions between the home ice effect and the REs in this study.

\begin{table}[ht]
\begin{center} 
\caption{ Yearly Coefficients for CORSI.
\label{tab:corsi:years}}

\begin{tabular}{l|r|r|r|r|r|r}
& \multicolumn{6}{c}{Season}\\
Effect&2007-08&2008-09&2009-10&2010-11&2011-12& 2012-13\\
\hline
mean, $\exp(\hat{\mu})$&	48.035	&49.818&	51.509&	52.630&	52.852&	52.018\\
ASD, $\exp(\hat{\beta})$ &0.969&	0.974&	0.963&	0.960&	0.964&	0.964\\
home ice, $\exp(\hat{\gamma})$& 1.070&	1.058&	1.061&	1.052&	1.058&	1.036
\end{tabular}
\end{center}
\end{table}

  \begin{table}[ht]
\begin{center} 
\caption{ Significant Rink Effects for CORSI Events.\label{tab:corsi:rinks}}
\begin{tabular}{l|r}
Rink & Effect\\
\hline
 TOR & 1.125\\
CAR&1.078\\
 MTL& 1.077\\
 S.J&1.053\\
EDM& 1.048\\
PHI & 1.030\\
  \hline
 FLA & 0.977\\
 CBJ& 0.964\\
 STL&0.963\\
  BOS& 0.951\\
 CHI&0.893\\
N.J & 0.839
\end{tabular}\\
There were no persistent 'homer' effects for CORSI.
\end{center}
\end{table}
 
 \subsection{Results for FENWICK}
 
 Another derived measure that is often used for statisical analysis is FENWICK.  
 Again, we apply the same models that we have used previously to study rink effects for this metric.
 FENWICK events are SHOTs, GOALs and MISSs.  As we saw previously with SHOTs and MISSs, the prorated number of FENWICKs per game has increased over this period, notably from a low of about $36.4$ events during the 2007-08 season to over $38$ events during the last four regular seasons considered here.  A summary of these results is found in Table \ref{tab:fenwick:years}. The score differential effect is relatively similar across all six years with an average rate of $0.975$ meaning that for each additional average goal of score differential the rate of FENWICKs decreases by a factor of about $2.5\%$.  The home ice effect for FENWICK declines slightly over the seasons analyzed.  On average the rate of FENWICKs is $1.06$ higher per game for home teams than for away teams.  
%
%
%
%
 
There were only five rinks that had significant REs for FENWICKs.   These REs are given in Table \ref{tab:fenwick:rinks}.  The significant rinks are Carolina (CAR), Chicago (CHI), Colorado (COL), New Jersey (N.J) and Toronto (TOR).  There were no significant rink by home team interactions for the data analyzed as part of this study.
 
\begin{table}[ht]
\begin{center} 
\caption{ Yearly Coefficients for FENWICK Events.\label{tab:fenwick:years}}

\begin{tabular}{l|r|r|r|r|r|r}
& \multicolumn{6}{c}{Season}\\
Effect&2007-08&2008-09&2009-10&2010-11&2011-12& 2012-13\\
\hline
mean, $\exp(\hat{\mu})$&	36.384	&37.565	&38.924
&39.969	&39.187	&38.018\\
ASD, $\exp(\hat{\beta})$ &	0.979&	0.987&	0.972&	0.974&	0.976&	0.977\\
home ice, $\exp(\hat{\gamma})$ &	1.075&	1.069&	1.066&	1.053&	1.060&	1.037
\end{tabular}
\end{center}
\end{table}

  \begin{table}[ht]
\begin{center} 
\caption{ Significant Rink Effects for FENWICK Events.\label{tab:fenwick:rinks}}
\begin{tabular}{l|r}
Rink & Effect\\
\hline
TOR& 1.074\\
CAR& 1.041\\
COL&1.017\\
\hline
N.J& 0.905\\
CHI& 0.893\\

\end{tabular}
\\
There were no persistent 'homer' effects for FENWICK.
\end{center}
\end{table}

 \subsection{Results for TURN}

The last metric we will consider in this section is TURN or turnovers.  
We credit a team with a TURN if they possess the puck after the event. The adjusted number of TURNs decreases slightly over the course of the seasons studied here.  Overall,
the expected number of TURNs is approximately $14.5$.  A summary of the results for our analyses can be found in Table \ref{tab:turn:years}.  The average score differential effect is around $1.04$ per goal of average score differential.  This effect seems to be relatively constant across the seasons.  The home ice effect here is at or just below $0.95$ and for the 2010-11 season there was not a significant home ice 
effect for TURNs.

  \begin{table}[ht]
\begin{center} 
\caption{ Yearly Coefficients for TURNs.
\label{tab:turn:years}}

\begin{tabular}{l|r|r|r|r|r|r}
& \multicolumn{6}{c}{Season}\\
Effect&2007-08&2008-09&2009-10&2010-11&2011-12& 2012-13\\
\hline
mean, $\exp(\hat{\mu})$&	15.096&	14.802&	15.157&	14.752&	13.852&	14.251\\
ASD, $\exp(\hat{\mu})$&
1.041	&1.046&	1.035&	1.039&	1.045&	1.033\\
home ice,$\exp(\hat{\gamma})$ &0.951&	0.945	&0.995&$^{\dagger}$	1.000&	0.955&	0.950
\end{tabular}\\
$^{\dagger}$ indicates that a coefficient was not significant.
\end{center}
\end{table}

  \begin{table}[ht]
\begin{center} 
\caption{ Significant Rink Effects for TURN Events.\label{tab:turn:rinks}}
\begin{tabular}{l|r}
Rink & Effect\\
\hline
 EDM&1.600\\
 NYI&1.548\\
 S.J&1.483\\
DAL&1.373\\
TOR&1.345\\
 OTT& 1.289\\
CGY&1.275\\
NSH&1.273\\
WSH&1.259\\
 COL&1.230\\
 L.A&1.186\\
\hline
 N.J&0.799\\
 NYR&0.777\\
 MIN&0.730\\
 STL& 0.693\\
 BOS&0.662\\
 FLA&0.671\\
 PIT&0.580\\
 CBJ&0.548\\
 PHX&0.455\\
\hline
VAN*HOME&0.901\\
PIT*HOME&0.832\\
N.J*HOME& 0.777\\

\end{tabular}
\end{center}
\end{table}

Table \ref{tab:turn:rinks} shows the rinks with significant REs for TURNs.  There are twenty such rinks.  
 There are significant home by rink interactions for TURNs at the following rinks: New Jersey (N.J), Pittsburgh (PIT), and Vancouver (VAN). All three of these rinks record TURNs by the home team at a rate that is significantly less than the rates in other rinks.
 There were also consistent TURN REs for both ATL and WPG in the same direction; however, because the current Winnipeg Jets has only been in their current building for two years, we cannot justify claiming that there is a persistent `homer' effect with this much data.  

\subsection{Summary of Rink Effects}
There are several overall trends worth highlighting in the above results.  First we note that there is a general trend for several rinks in particular. 
We will focus on SHOTs, MISSs, HITs and BLOCKs here since they are most commonly used by hockey analysts.  
It is fairly clear that events at the Prudential Center home of the New Jersey Devils is a place where events are recorded at a rate that is much lower compared to the rest of the league.  That rink has persistent effects for all four of the events considered here.  Though not as drastic as is the case for the New Jersey, Edmonton, Los Angeles and Toronto all have persistent effects for MISSs, HITs and BLOCKs.  Florida and St. Louis have persistent
REs for the counting of SHOTs though these
effects are small relative to many of the
others considered here.  
%
%
Above, we have given results for individual aspects of our analysis.  Here we want to offer some overall summaries.

\begin{figure}
\centering
\textbf{\Large Eastern Conference}
\subfloat[ Atlantic Division]{\includegraphics[width = 6.5in]{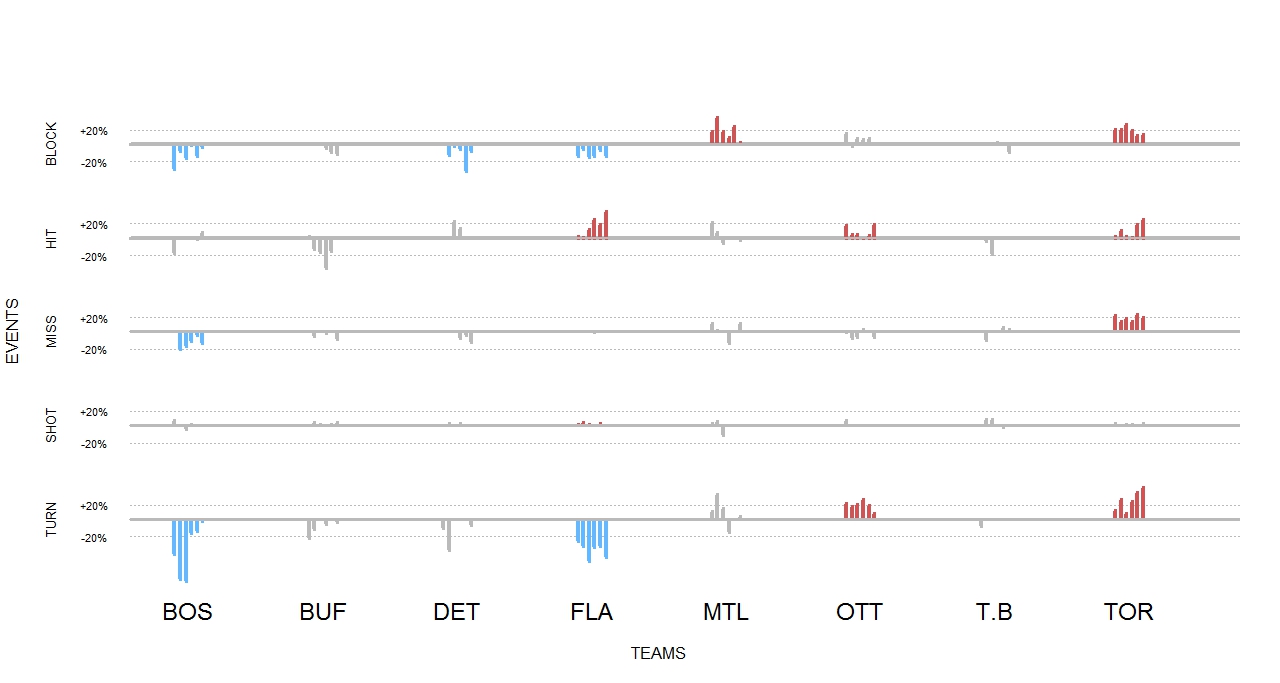}}\\ 
\subfloat[ Metropolitan Division]{\includegraphics[width = 6.5in]{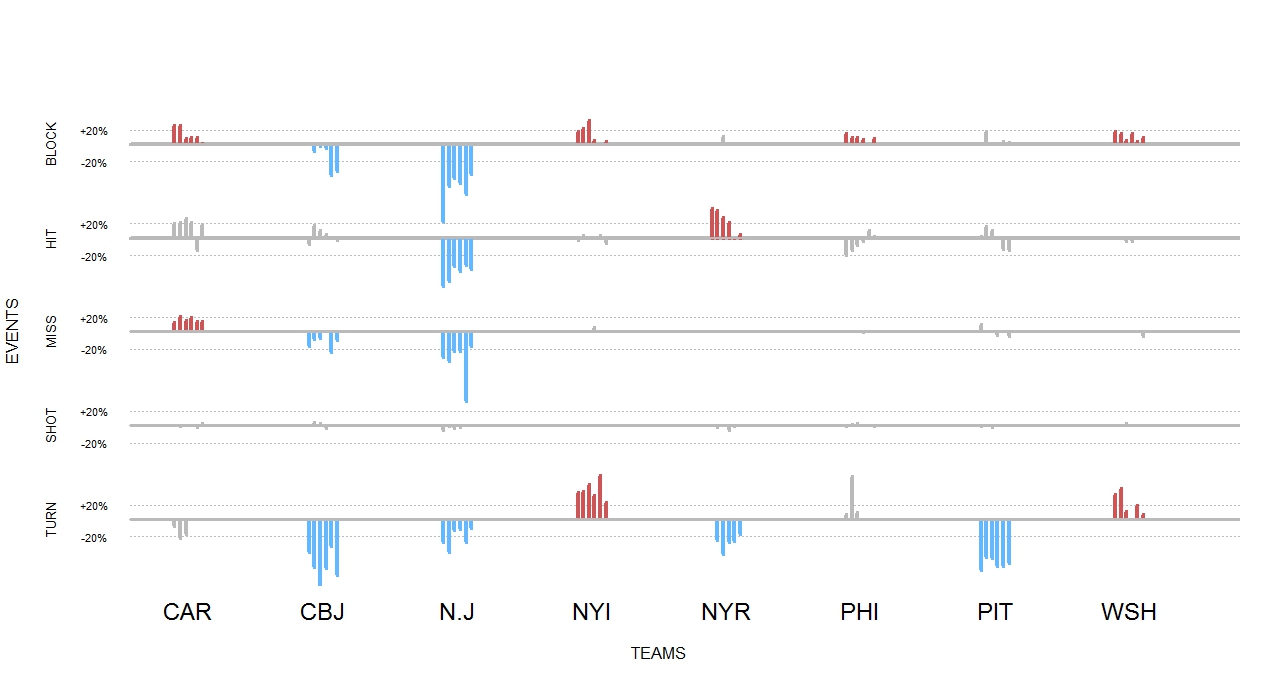}}
\caption{Eastern Conference Rink Effects by Division for some NHL events. Red bars indicates that for that event a particular rink persistently recorded events above the average in all rinks, while blue bars indicates that a rink persistently recorded events below the average of other rinks.  The events listed are TURN, SHOT, MISS, HIT, BLOCK.  Results for Winnipeg (WPG) are from both Atlanta and Winnipeg.\label{bigplot_east1}}
\end{figure}

\begin{figure}
\centering
\textbf{\Large Western Conference}
\subfloat[ Central Division]{\includegraphics[width = 6.5in]{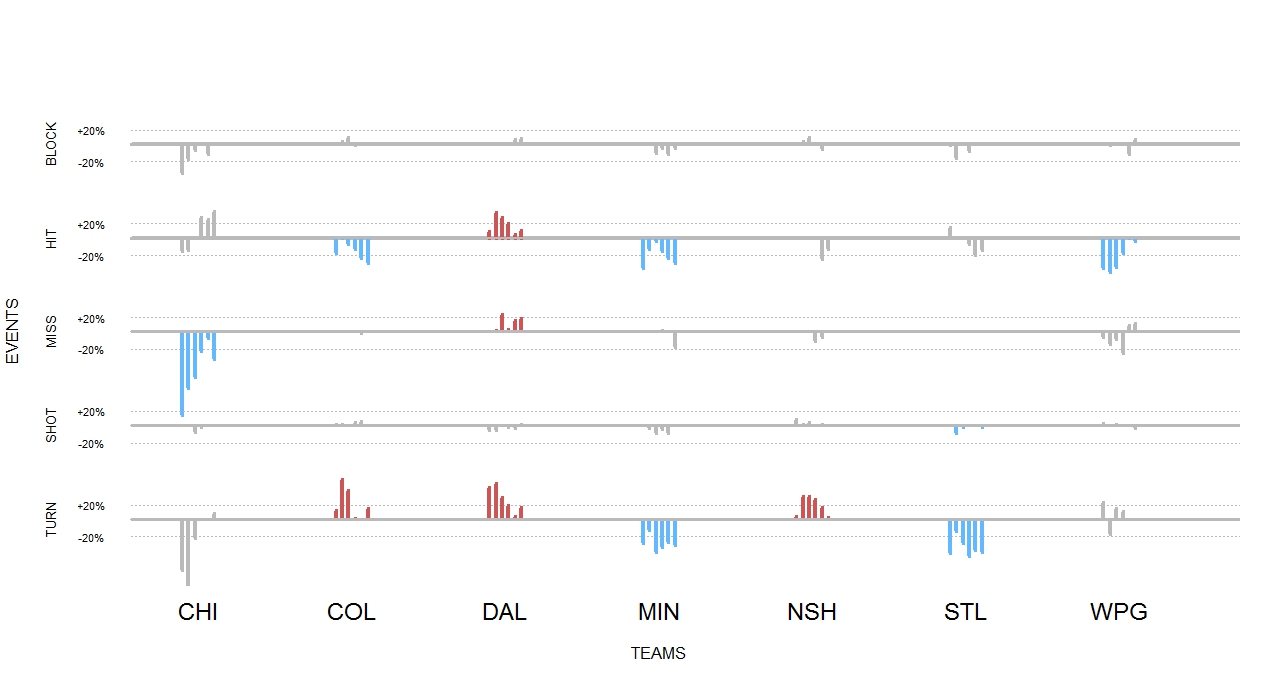}}\\ 
\subfloat[ Pacific Division]{\includegraphics[width = 6.5in]{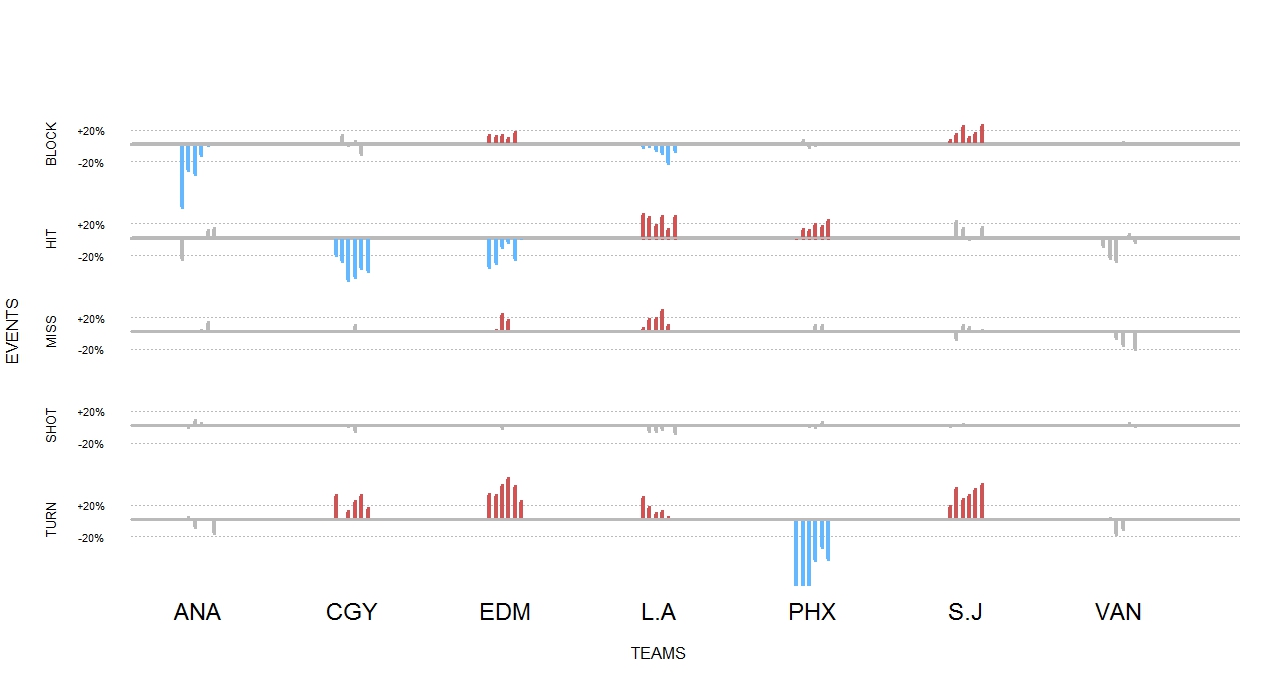}}
\caption{Western Conference Rink Effects for Western Conference by Division. Red bars indicates that for that event a particular rink persistently recorded events above the average in all rinks, while blue bars indicates that a rink persistently recorded events below the average of other rinks.   The events listed are SHOT, MISS, BLOCK, CORSI, FENWICK.  Results for Winnipeg (WPG) are from both Atlanta and Winnipeg.\label{bigplot_west1}}
\end{figure}

\begin{figure}
\centering
\textbf{\Large Eastern Conference}
\subfloat[ Atlantic Division]{\includegraphics[width = 6.5in]{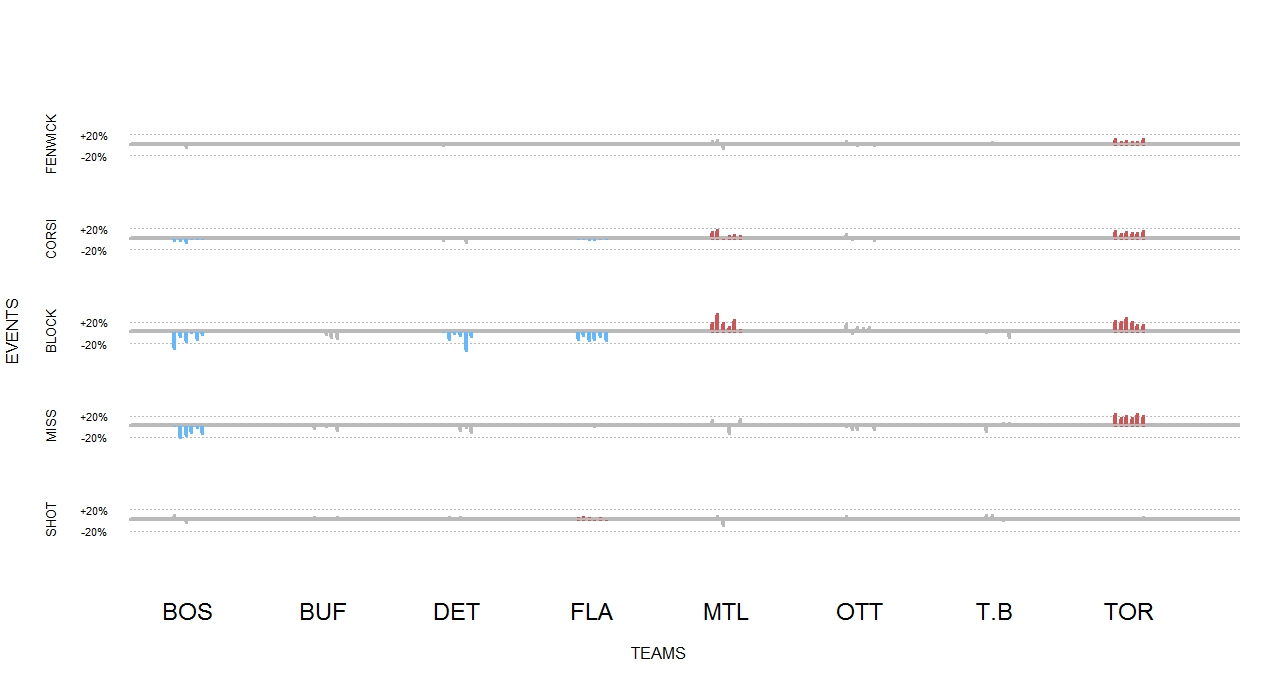}}\\ 
\subfloat[ Metropolitan Division]{\includegraphics[width = 6.5in]{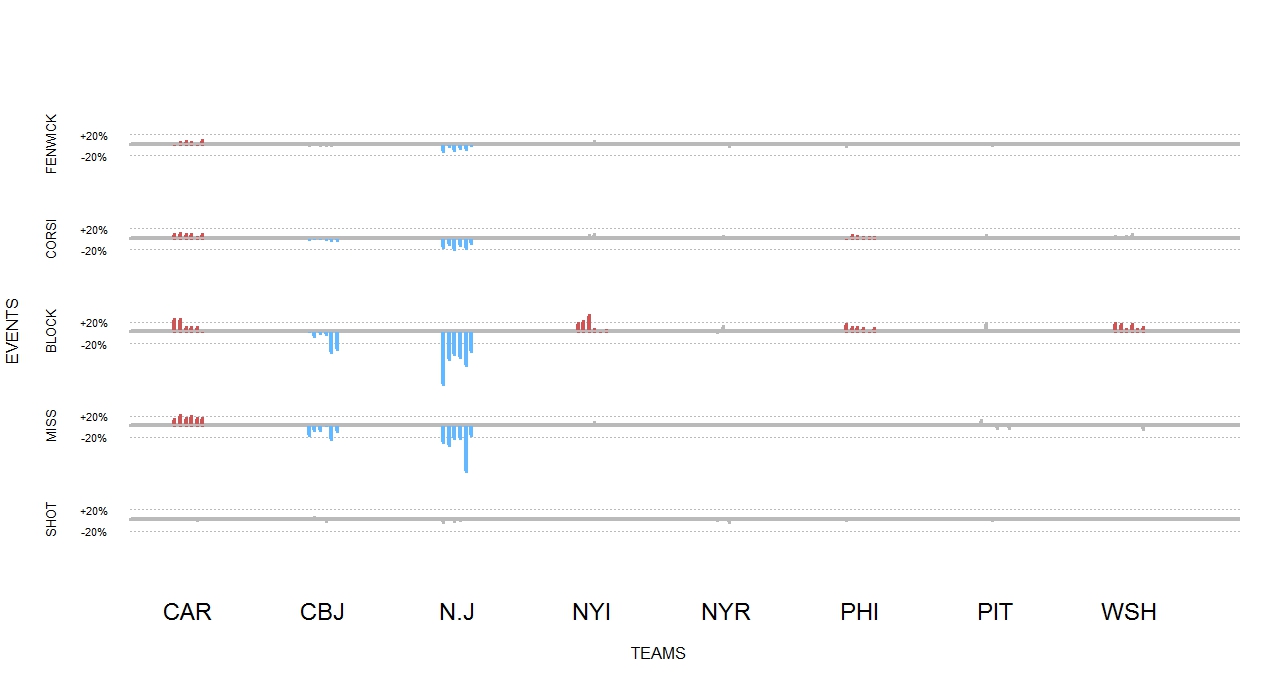}}
\caption{Rink Effects for Eastern Conference by Division. Red bars indicates that for that event a particular rink persistently recorded events above the average in all rinks, while blue bars indicates that a rink persistently recorded events below the average of other rinks. The events listed are SHOT, MISS, BLOCK, CORSI, FENWICK.  Results for Winnipeg (WPG) are from both Atlanta and Winnipeg.\label{bigplot_east2}}
\end{figure}

\begin{figure}
\centering
\textbf{\Large Western Conference}
\subfloat[ Central Division]{\includegraphics[width = 6.5in]{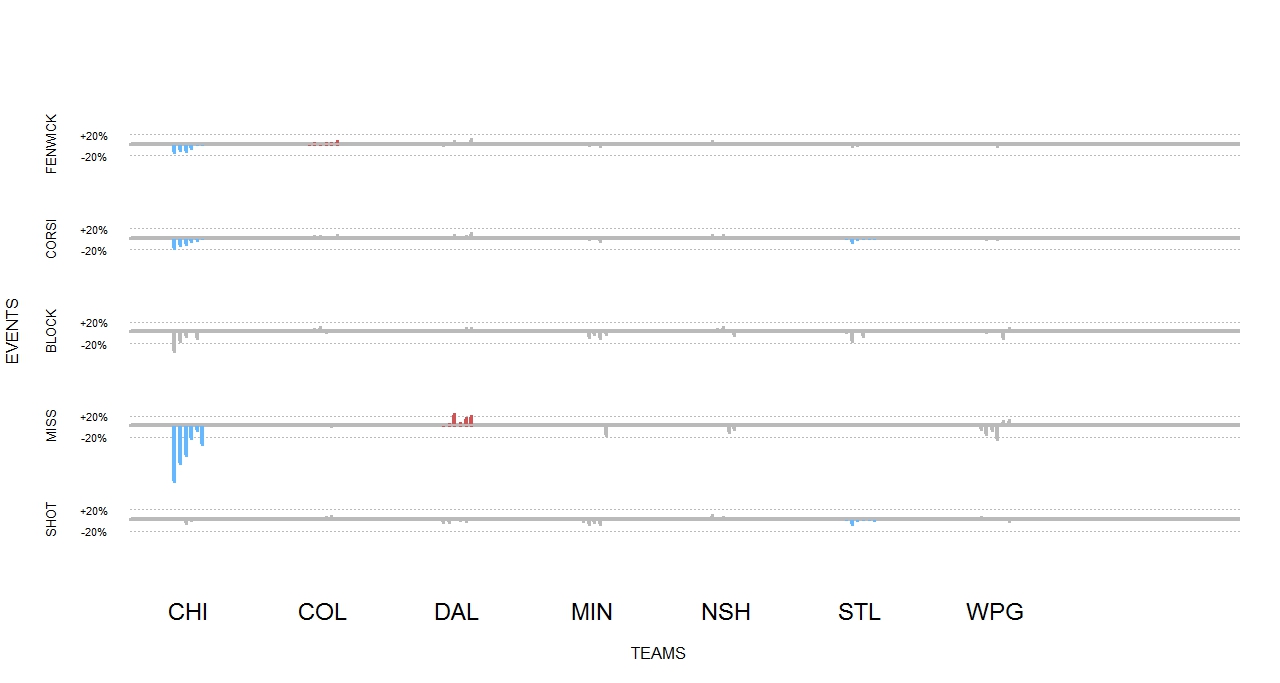}}\\ 
\subfloat[ Pacific Division]{\includegraphics[width = 6.5in]{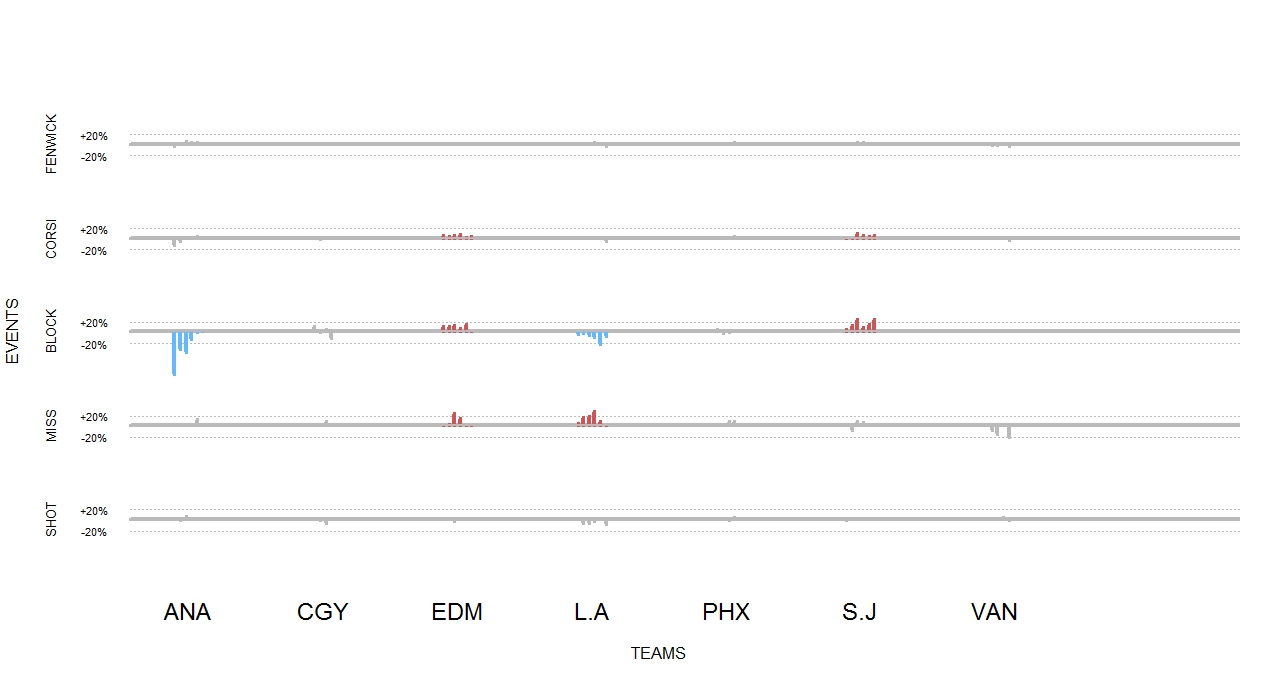}}
\caption{Rink Effects for Western Conference by Division. Red bars indicates that for that event a particular rink persistently recorded events above the average in all rinks, while blue bars indicates that a rink persistently recorded events below the average of other rinks.  The events listed are SHOT, MISS, BLOCK, CORSI, FENWICK.  Results for Winnipeg (WPG) are from both Atlanta and Winnipeg.\label{bigplot_west2}}
\end{figure}

Table \ref{tab:overall:rinks} has a summary of the results from this section.  Plots of the rink effects for some events,
by year, conference and division can be
found in Figures \ref{bigplot_east1},
\ref{bigplot_west1}, \ref{bigplot_east2}, and
\ref{bigplot_west2}.  
For each event we
give the number of rinks with persistent
RE's, the number of rinks with persistent `homer' effects (HE's) and the range of 
RE's for that event.  From that table we 
can see that there are double digit 
persistent rink effects for all of the 
events considered here outside of MISSs, 
SHOTs and FENWICK.  Persistent `homer' 
effects have been noted for HITs, MISSs 
and TURNs.  The largest range of effects 
is found for GIVE and TAKE which is not 
surprising given the results found in 
previous analyses.  

Figures \ref{bigplot_east1} and \ref{bigplot_west1} have summaries of REs for BLOCKs, HITs, MISSs, SHOTs and TURNs while Figures \ref{bigplot_east2} and \ref{bigplot_west2} have summaries of REs for FENWICKs, CORSIs, BLOCKs, MISSs and SHOTs.  For those graphs each vertical bar on the graphs represents the REs for a particular rink and a given year.  For those rinks where the REs are persistent we highlight them in red for above average and blue for below average. The rinks --- Edmonton, Los Angeles, New Jersey and Toronto --- which have persistent effects for BLOCKs, HITs and MISSs can clearly be seen in these graphs.  Further, the large amount of variation in the RE's for TURN are clear from these graphs.

We note that SHOTs, which is arguably the most important event that is recorded by the NHL's RTSS system, only has two persistent HEs, has the smallest range of REs, and has relatively small average ASD and home ice effects.  This is good news
as SHOTs are an integral component of many important metrics of hockey performance.  
%
%

The effect of score differential, ASD, across years is relatively close to one ranging only from 0.92 to 1.06.  Home ice, however, has a much larger range, mostly due to GIVE and TAKE.  TURNs have a below average rate for home teams relative to away teams and they are the only event with that characteristic.

 \begin{table}[ht]
\begin{center} 
\caption{ Comparison of RE's across Events.\label{tab:overall:rinks}}
~\\
\begin{tabular}{l|r|r|r|r|r}
&Number of&Number of & Range of&Average&Average\\
Event&Persistent RE's&Persistent HE's&RE's&ASD effect&home ice effect\\
\hline
BLOCK&15&0&0.73&0.92&1.05\\
GIVE&18&0&2.02&0.95&1.68\\
HIT&12&3&0.71&0.96&1.11\\
MISS&9&2&0.69&0.95&1.12\\
SHOT&2&0&0.08&0.96&1.05\\
TAKE&17&0&1.73&1.06&1.54\\
CORSI&12&0&0.29&0.96&1.06\\
FENWICK&5&0&0.18&0.98&1.06\\
TURN&20&3&1.14&1.04&0.97\\
\end{tabular}
\end{center}
\end{table}

The average effect here for home ice is much smaller than those observed for GIVE and TAKE; however, there are slightly more rinks with persistent REs for TURN and there are persistent home by rink interactions or home effects for TURN while there are none for GIVE and TAKE.  
%
%

  \begin{table}[ht]
\begin{center} 
\caption{ Adjusted HIT Counts for 2012-13 NHL Regular Season \label{tab:hits:revised1213}}
~\\
\begin{tabular}{l|c|r|r|r}
Name&Team&Adjusted Hits&Raw Hits&Differential\\
\hline\hline
M Martin &NYI&228.9&234&-5.1\\
C Neil &OTT&193.4&206&-12.6\\
S Ott &BUF&182.6&187&-4.4\\
L Schenn &PHI&181.8&187&-5.2\\
C Clutterbuck &MIN&181.3&155&26.3\\
L Smid &EDM&173.0&151&22.0\\
L Komarov &TOR&169.7&176&-6.3\\
D Backes &STL&159.9&158&1.9\\
R Clune &NSH&159.0&159&0.0\\
M Fraser &TOR&147.7&153&-5.3\\
E Kane &WPG&143.0&147&-4.0\\
Z Rinaldo &PHI&140.4&143&-2.6\\
M Lucic &BOS&137.4&139&-1.6\\
R Reaves &STL&137.0&135&2.0\\
D Brown &LAK&136.4&156&-19.6\\
K Clifford &LAK&135.8&155&-19.2\\
R Callahan &NYR&132.0&154&-22.0\\
J Petry &EDM&128.2&112&16.2\\
B Boyle &NYR&126.9&145&-18.1\\
D MacKenzie &CBJ&124.2&122&2.2\\

\end{tabular}
\end{center}
\end{table}

 \section{Discussion}

In this paper we have developed a flexible methodology for estimation of the rink effects for the statistics used by the National Hockey League.  Using data from six NHL seasons we have estimated which rinks have persistent differences in their rates of recording events.  We have considered both primary events, for example, SHOTs, but also derived events including turnovers (TURN) and CORSI events.  Several facets of our model are noteworthy.  First, we have developed an elastic net regression methodology that accounts for team effects, score effects, and home ice in order to isolate the impact of a particular rink.  The elastic net forces to zero terms that are not significant.
Second, we have introduced average score effects for NHL games which look at the time-averaged score differential over the course of a game.  From an analysis of the residuals from our model a log-linear average score differential seems to be appropriate for the data analyzed as part of this paper.  Third, we use the natural
logarithm of the event counts which leads to a multiplicative model.  That is, the effects (rink, home ice, etc.) impact the counts through a multiplication of these effects.  

Next we illustrate the impact of these rink effects by looking at how these effects impact counts for individual players.  Applying our rink effects model to each HIT from the 2012-13 NHL season, we can get the adjusted HIT count for each player.  Table \ref{tab:hits:revised1213} contains the players with the top 20  adjusted HIT counts for that season.  To get those adjustments for rink $j$, we take each count and weight it as $1/\hat{\gamma}_j$  if there was not a 'homer' effect and by $1/(\hat{\gamma}_j \hat{\eta \gamma})_j$ if there is such an effect. For example each hit that occurs in the Staples Center, home of the Los Angeles Kings, should be weighted as $1/1.297=0.77$ or about 77\% of the weight given to hits recorded in rinks without significant REs.  Hits recorded in the Prudential Center, home of the New Jersey Devils, for the away team should be weighted as $1/0.581$ or about 172\% of hits in a normal rink.  For the Devils at home, HITs should be weighted as $1/(0.581\times1.186)=1.45$ or about 145\% of hits in normal rinks.  As a consequence of these weightings we can see that hit counts for players in arenas with large RE's had substantial changes to their hit count.  Cal Clutterbuck moves from a count of 155 to an adjusted count of 180.9 due to the relatively lower rate of recording of hits in his home rink, Xcel Energy Center.  While Clutterbuck moves up, Dustin Brown, Kyle Clifford and Ryan Callahan have much lower adjusted HITs due to the rinks in which they play.

 \begin{table}[ht]
\begin{center} 
\caption{ Adjusted BLOCK Counts for 2012-13 NHL Regular Season \label{tab:blocks:revised0910}}
~\\
\begin{tabular}{l|c|r|r|r}
Name&Team&Adjusted Blocks&Raw Blocks&Differential\\
\hline\hline
F Beauchemin &ANA&133.3&111&22.3\\
 G Zanon &COL&127.2&124&3.2\\
 D Girardi & NYR&120.7&125&-4.3\\
 R Hainsey &WPG&120.5&123&-2.5\\
 D Seidenberg & BOS&119.8&115&4.8\\
 L Smid &EDM&114.7&119&-4.3\\
 B Orpik &PIT&110.3&114&-3.7\\
 A MacDonald &NYI&109.8&123&-13.2\\
 J Carlson &WSH&109.4&123&-13.6\\
 N Schultz &EDM&106.8&107&-0.2\\
 J Hejda &COL&105.7&102&3.7\\
 A Pietrangelo &STL&104.9&101&3.9\\
 B Seabrook &CHI&104.1&103&1.1\\
D Wideman &CGY&101.9&100&1.9\\
 J Gorges &MTL&100.4&116&-15.6\\
 J Johnson & CBJ&100.2&90&10.2\\
D Murray &S.J, PIT&100.1&98&2.1\\
 T Hamonic &NYI&100.1&109&-8.9\\
 J Harrison &CAR&96.6&110&-13.4\\
 N Hjalmarsson &CHI&96.2&94&2.2\\
\end{tabular}
\end{center}
\end{table}

Another example of the application of our estimated Rink Effects can be found in Table \ref{tab:blocks:revised0910}.  Unlike in Table \ref{tab:hits:revised1213} where there was not a change at the top for adjusted HITs, here the top player on the revised list, Francois Beauchemin was originally ranked $10^{th}$ on the list of players with the most blocks.  There are other large changes for players from the Islanders, Andrew MacDonald, and the Capitals, John Carlson, due to the relative counting of BLOCKs in those rinks.  MacDonald and Carlson move from tied for third in raw BLOCKs to eighth and ninth, respectively, after our adjustment.  Similarly, Josh Gorges moves from seventh to fifteenth.

We now illustrate the impact of rink adjustments on CORSI  percentages.  As above, for this adjustment we reweight each even strength CORSI event by the RE's for each rink using the rink effects in Table \ref{tab:corsi:rinks}.  Thus we count 
each CORSI event in Toronto as worth $1/1.125=0.89$ of an event while a CORSI event in New Jersey is counted as worth $1/0.839=1.19$ of an event.  These adjustments are done for all rinks with persistent effects for data from the 2012-13 regular season.  
We report the percent of all CORSI events `for' a given team from all of the CORSI events that occurred when they were involved for the top five and bottom five teams in Table \ref{tab:corsi:revised}.  Note that many authors have used this percentage as a measure of team quality.  
From that table,  it is clear that CORSI rink effect adjustments have little impact on the CORSI percent.  (We have had to take these proportions out to the fourth decimal place to show that differences do result from these adjustments.)  Largely this is due to the fact that the impact of the rink effects is felt both the counts `for' and `against' a given team.  Thus, both the numerator and the denominator are affected by these CORSI rink effects and the impacts balance each other out.  That is why the adjusted ratios given in Table \ref{tab:corsi:revised} are similar to the raw ratios.
%
%
In a similar vein, we used team for and team against FENWICK ratings from our model to predict Stanley Cup playoff matches and found that these ratings predicted winners from these matches at the same rate as score adjusted Fenwick rates created by Tulsky \cite{tulsky_fenwick}.  

 \begin{table}[ht]
\begin{center} 
\caption{ Adjusted CORSI Percentage for 2012-13 NHL Regular Season for Top 5 and Bottom 5 Teams \label{tab:corsi:revised}}
~\\
\begin{tabular}{l|r|r}
Team&CORSI Pct.&CORSI Pct.\\
&&adjusted\\
\hline
L.A   &     0.5630 &          0.5628\\
N.J   &     0.5592 &          0.5592\\
BOS   &     0.5433 &          0.5430\\
CHI   &     0.5414 &          0.5420\\
DET   &     0.5366 &          0.5365\\
\hline
CBJ   &     0.4711  &         0.4706\\
NSH   &     0.4668  &         0.4662\\
BUF   &     0.4512  &         0.4513\\
EDM   &     0.4458  &         0.4445\\
TOR   &     0.4408  &         0.4398\\
\end{tabular}
\end{center}
\end{table}

From the analyses above,  it is clear that there are differences in the recording of the NHL's RTSS
Events among rinks in the NHL.  Event counts can see noticeable differences as
illustrated above with BLOCKs and HITs.  These rink effects (and their adjustments) 
impact statistics that are used in the player arbitration process since the RTSS quantities are
eligible to be included during that process, Hancock \cite{cba}.  Percentages, such as CORSI
percentages, are much less affected by the rink effects that we have estimated here. 
Our estimation of these rink effects for the counts of events is based upon 
a flexible log-linear regression model
that accounts for the teams involved, for the average score differential, for home ice 
and for a home ice by rink interaction.  We applied these models to the six NHL 
regular seasons from 2007-08 to 2012-13.  Our focus has been the rink effects and 
adjustments derived from these estimated rink effects.    
For BLOCKs, HITs, and MISSs, there were six 
rinks: Buffalo, Nashville, Pittsburgh, St. Louis, Tampa Bay and Vancouver, who had no persistent effects.  Four rinks: Edmonton, Los Angeles, New Jersey and Toronto had persistent effects for all three events.  And two rinks had persistent effects for SHOTs: Florida and St. Louis.


%


%
%
%
%

\newpage
\bibliography{bibsports.bib}
\bibliographystyle{unsrt}

\end{document}